# Nonadiabatic couplings drive ultrafast, mode-selective intramolecular vibrational energy redistribution in flavins

Daniel Timmer,[a+] Krishan Kumar,[a+] Jan P. Götze,[b] Peter Saalfrank,[c] Antonietta De Sio,[a,d] and Christoph Lienau[a,d*]

[a] Institut für Physik, Carl von Ossietzky Universität Oldenburg, 26129 Oldenburg, Germany

[b] Institut für Chemie und Biochemie, Freie Universität Berlin, 14195 Berlin, Germany

[c] Institute of Chemistry, University of Potsdam, 14476 Potsdam, Germany

[d] Center for Nanoscale Dynamics (CENAD), Carl von Ossietzky Universität Oldenburg, 26129 Oldenburg, Germany

*Correspondence to: christoph.lienau@uni-oldenbug.de

+Both authors contributed equally.

## Abstract

Flavins are the chromophores in several blue-light-sensitive photoreceptor proteins and act as redox cofactors in many enzymes relevant for biological processes. Despite their biological relevance and numerous, detailed optical investigations of their photophysical properties, the ultrafast nonequilibrium dynamics of their elementary optical excitations are not yet fully known. Here, we use ultrafast coherent vibrational spectroscopy with 10-fs time resolution in the 450-nm spectral range to study their excited state coherent vibrational dynamics. We observe that coherent wavepacket motion along high-frequency C-C stretching modes with ~ 20-fs period is rapidly damped on a 20-fs timescale. In contrast, coherent motion along several low-frequency modes persists much longer. We attribute this to a mode-selective intramolecular vibrational energy redistribution driven by nonadiabatic couplings between the optical bright state and a close-lying dark electronic state, in accordance with model calculations. Our results may be of relevance for the formation of long-lived radical pair states in magnetic-field sensitive proteins.

**Introduction**

Flavins are prototypical blue-light-absorbing chromophores acting as cofactors for redox-active enzymes in various biological systems. Flavoproteins play a crucial biological function, such as DNA repair,[1,2] regulation of circadian rhythm in plants[3] and animals,[4,5] as well as functions related to plant growth.[6,7] Their isoalloxazine moiety, central to flavin-based cofactors such as riboflavin, flavin mononucleotide, and flavin adenine dinucleotide (FAD), is primarily responsible for their photophysical properties. FAD is a widespread cofactor found in several flavoproteins, such as the class of BLUF (blue light sensing using FAD) proteins, flavin-containing monooxygenase, LOV (light, oxygen, voltage) proteins, and, in particular, blue-light-absorbing cryptochrome photoreceptors.[8-13] Cryptochromes (Crys) fulfill a wide range of functions in plants and animals. Photoexcitation of cryptochrome triggers a series of electron transfer steps from nearby amino acid residues, leading to the formation of long-lived, spin-correlated radical-pairs.[14-17] Singlet-triplet interconversion of this radical pair is currently considered the main candidate for the microscopic origin of the magnetic field sense in migratory song birds.[15, 18-20] Indeed, such magnetic field sensitivity has been shown *in vitro* for FAD-binding European Robin cryptochrome 4 (*Er*Cry4).[21] Furthermore, flavin-based molecular chromophores are also actively studied for photocatalytic transformation of small organic molecules, e.g., photooxidation of alcohols,[22, 23] cyclization of barbituric acid derivatives[24] and amide bond formation between aliphatic or aromatic aldehydes and amines.[25]

Owing to their broad functional significance, the photophysical properties of flavins and flavoproteins have been extensively investigated.[16, 26-34] In particular, time-resolved experiments such as transient absorption (TA) and time-resolved fluorescence spectroscopy have been used to investigate the photophysical fate of the excited state. Time-resolved fluorescence spectroscopy has shown that the fluorescence properties of FAD vary between open and stacked conformations of the adenine and isoalloxazine rings, with strongly quenched emission from stacked conformation.[31, 35-37] The strong emission quenching has been largely[32] assigned to an intramolecular electron transfer,[31, 38-40] and the influence of pH on the excited state of the two conformers has been investigated using TA.[41, 42] TA has also been used to explore the dynamics of the different redox states of FAD[21, 28] that exist in planar or bent isoalloxazine ring structure, with the latter structure resulting in faster deactivation of the excited state.[16, 31] Furthermore, stimulated Raman spectroscopy has been used to investigate the vibrational modes of FAD in the excited state without interference from ground state modes of FAD in polar solvents.[33] This study also points to a central role of vibronic couplings for intramolecular vibrational energy redistribution (IVR) in the optically excited states.

Flavins show a broad absorption spectrum in the blue (Fig. 1b) with two apparent electronic transitions, a lower-energy band close to 450 nm and an additional band in the UV close to 370 nm. Theoretical studies of the electronic structure[43-48] assign these bands to the two lowest-energy $\pi\pi^*$ transitions in the molecule. While the 450-nm band is commonly associated with the $S_0 \rightarrow S_1$ transition, different labeling of the higher-lying electronic states are found in the literature. We follow Ref. 47 and associate the 370-nm band with the $S_0 \rightarrow S_4$ transition (see Table 1 in Ref. 47) Electronic structure calculations give evidence for two additional, optically dark $n\pi^*$ transitions to states $S_2$ and $S_3$ in this energy region. The energy spacing between $S_1$ and the closest, higher-lying $S_2$ state is strongly solvent dependent.[33, 49] Mixed quantum classical calculations predict that nonadiabatic couplings between $S_1$ and $S_2$ play a fundamental role for the excited state dynamics in flavins. They suggest that such couplings are driven by high-frequency (~22 fs) excited state vibrations.[47] They induce a strong red-shift of the stimulated emission spectra within the first few 10 fs after excitation and result in rapid IVR processes which dampen the high-frequency vibrational wavepacket motion on a sub-100 fs time scale.[47] Recently, related nonadiabatic dynamics have been discussed theoretically in cryptochromes[50, 51] and low-lying conical intersections were predicted in time-dependent density functional theory (TD-DFT) simulations of lumiflavin.[52] So far, direct time-domain studies of the role of such nonadiabatic couplings on the excited state dynamics of flavin systems are lacking. Quite generally, experimental studies of nonadiabatic dynamics in solution and in thin films with a time resolution faster than the period of the relevant vibrational modes are limited.[53-58]

Here, we report on the experimental observation of nonadiabatic couplings in FAD by directly following the excited state dynamics using TA spectroscopy with high time-resolution. Utilizing 12-fs blue excitation pulses, we observe that the vertically excited, Franck Condon state decays within ~20 fs accompanied by the ultrafast formation of a red-shifted stimulated emission band. Our results suggest that nonadiabatic couplings result in the rapid loss of coherent high-frequency vibrational motion in the excited state, while oscillations resulting from low-frequency modes persist on picosecond timescale. Our results reveal how vibronic coupling shapes excited state dynamics in a key biological cofactor, with implications for understanding light-driven function in flavoproteins.

**Results and Discussion**

As in all flavins, the optically relevant moiety in FAD is an isoalloxazine motif with its center ring connected to an adenine dinucleotide unit through a ribityl chain (Fig. 1a). The steady state absorption spectrum of an aqueous solution of FAD (Fig. 1b, black line) shows the typical flavin absorption in the blue spectral region close to 2.75 eV (450 nm) that can be assigned to the $S_0 \to S_1$ electronic transition of the isoalloxazine moiety, primarily involving excitation from a $\pi_2$ highest occupied molecular orbital (HOMO) to the $\pi_3$ lowest unoccupied molecular orbital (LUMO).[48] A second absorption band close to 3.35 eV (370 nm) has been assigned to a transition from $S_0 \to S_4$ involving excitation from HOMO-1 to LUMO.[48, 59-62] Theoretical calculations show that, in addition to these two optically bright excited states, two dark states $S_2$ and $S_3$ with $n\pi^*$ character lie energetically close to $S_1$.[43-46] The $S_0 \to S_1$ absorption band in Fig. 1b exhibits three faint shoulders close to 2.59 eV (~480 nm), 2.75 eV (~450 nm), and 2.91 eV (~426 nm). We assign these shoulders to $|S_0, 0\rangle \to |S_1, n\rangle$ vibronic transitions, where $n = 0, 1, 2$ denotes the quantum number of a high-frequency, ~160 meV (~1290 cm$^{-1}$) vibrational mode (Fig. 1b, stick spectrum). This vibronic progression is more pronounced in a rigid protein environment than in aqueous solution.[14] We also note that the vibrational fine-structure is in excellent agreement with a computed absorption spectrum.[59] The experimental emission spectrum of FAD in water (Fig. 1b, red line) shows an essentially mirrored, red-shifted emission band centered around 2.25 eV (~550 nm).

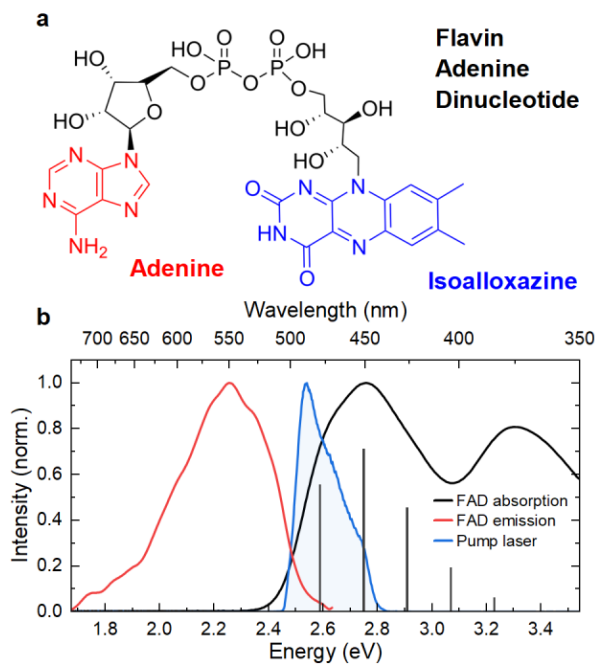

**Figure 1: (a)** Chemical structure of flavin adenine dinucleotide (FAD), consisting of a blue-light-absorbing isoalloxazine moiety that is connected to adenine via a ribityl chain. (b) Normalized steady-state absorption spectrum of FAD in water (black), showing the dominant $S_0 \rightarrow S_1$ transition at ~ 2.8 eV and the $S_0 \rightarrow S_4$ transition at ~ 3.3 eV. A faint vibronic progression with a dominant 160 meV vibrational mode is observed in the lower-energy band, as indicated by a stick spectrum. The red-shifted emission spectrum is shown in red. The spectrum of the pump pulses (blue) is chosen to overlap with the two vibronic resonances of FAD in the $S_1$ electronic state.

## Ultrafast TA spectroscopy of FAD

To experimentally investigate the role of coherent vibrational dynamics and, in particular, the effects of nonadiabatic couplings on these dynamics, we perform transient absorption (TA) spectroscopy of a 200 µM aqueous solution of FAD. To generate ultrashort and broadband excitation pulses, we developed a 10-kHz TA setup based on supercontinuum pulses generated in a Neon-filled hollow-core fiber. While the broadband pulses[63] are directly used as the probe, a 4f spectral filter is employed to prepare tunable pump pulses that are centered around 470 nm (Fig.

1b, blue line). These pulses are compressed using chirped mirrors to a pulse duration of 12 fs as measured via transient grating frequency-resolved optical gating (see Methods). The pulse compression substantially improves on the time resolution reached in earlier studies of flavins,[2, 16, 30-32, 40, 49, 64-66] including our own work based on an optical parametric amplifier.[14, 67] The aim is to resolve coherent oscillations induced by impulsive excitation in the electronic ground and excited states of all expected Raman-active modes of FAD, including high-frequency C-C stretching modes in the ~1500 cm$^{-1}$ range. The employed pump pulse mainly overlaps with the ground and the first excited vibronic state in the $S_0 \rightarrow S_1$ transition of FAD (see Fig. 1b). Due to the ultrashort nature of the pump pulses and low molar extinction coefficient of FAD (11300 M$^{-1}$cm$^{-1}$),[68, 69] the TA data exhibit significant cross-phase modulation (XPM) during pulse overlap at zero delay.[70] To minimize XPM contributions and uncover dynamics for short delays,[71, 72] the broadband probe is also compressed with chirped mirrors, and a reference measurement of the bare solvent is recorded under identical experimental conditions. The obtained $\Delta T/T$ map of FAD data is presented in Fig. 2a after careful chirp correction and solvent subtraction (Supporting Information, section 3). The TA map shows a ground state bleaching (GSB) band centered around 2.75 eV, a stimulated emission (SE) band centered around 2.25 eV, and three excited state absorption (ESA) bands centered around 2.45 eV, above 3 eV, and below 2 eV, as labelled in Fig. 2a. This assignment of the bands is well established in the literature.[33, 73]

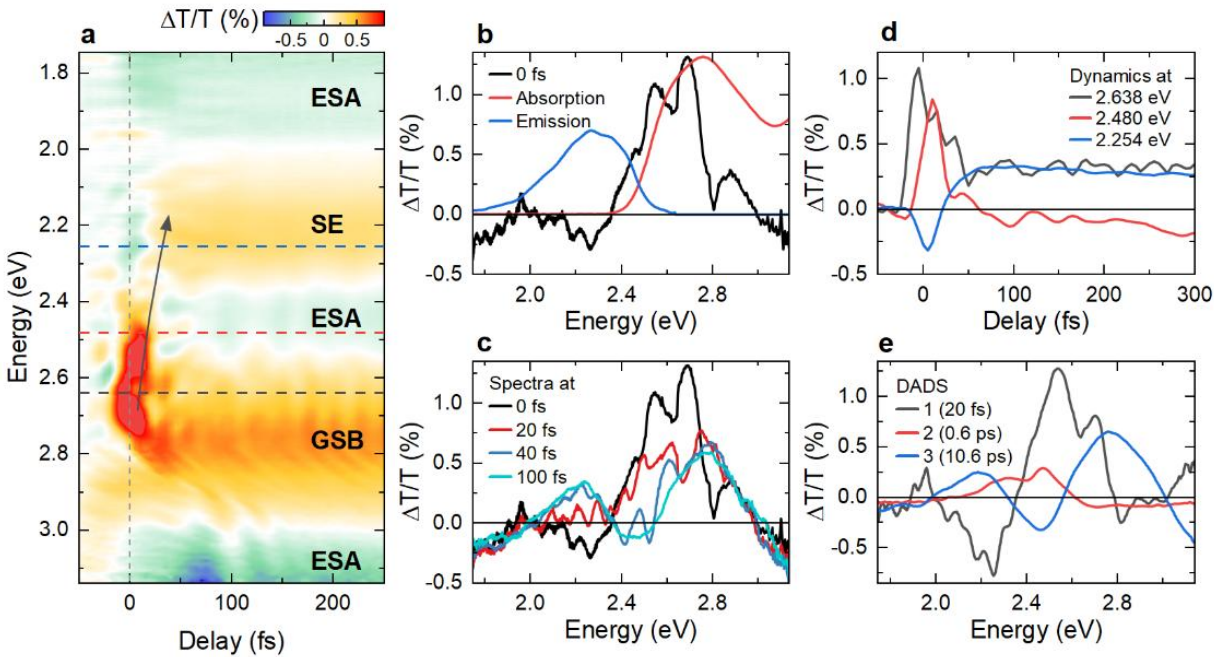

**Figure 2:** Transient absorption spectroscopy of FAD dissolved in water. (a) Map of differential transmission $\Delta T/T$ spectra of FAD as a function of pump-probe delay and probe energy. The 12-fs pump pulses are centered around 2.6 eV (Fig. 1). Important regions in the map are associated with ground-state bleaching (GSB), stimulated emission (SE), and excited-state absorption (ESA) transitions. SE from the Franck-Condon region, initially centered around 2.6 eV, displays a rapid red-shift during the first 50 fs (black arrow). This red-shift is taken as the signature of rapid intramolecular vibrational energy redistribution (IVR) in FAD. (b) Comparison of $\Delta T/T$ spectrum at $t = 0$ fs (black) with scaled steady-state absorption (red) and emission (blue) spectrum from Fig. 1. (c) $\Delta T/T$ spectra at selected delays, highlighting the transient red-shift of the SE and the rapid build-up of a relaxed SE band centered around 2.2 eV. (d) $\Delta T/T$ dynamics at selected probe energies reveal the rapid decay of SE from the Franck-Condon region (black), the transient red-shift (red), and the buildup of the relaxed SE (blue). (e) Decay-associated differential spectra (DADS) obtained by a global analysis. The corresponding decay times are given in the legend.

The $\Delta T/T$ spectrum at t = 0 fs, recorded during impulsive optical excitation, is depicted as a black line in Fig. 2b. The spectrum is centered around 2.6 eV, and its line shape is somewhat similar to the steady state absorption spectrum of FAD in water (red line). The spectrum displays a peak splitting of ~140 meV, similar to the energy of the high-frequency vibrational mode that is coupled to the electronic excitation. The peak energies are slightly red-shifted by ~70 meV with respect to those deduced from the absorption spectrum. We therefore tentatively assign the two dominant peaks at 2.55 eV and to 2.69 eV to the $|S_1, 0\rangle \to |S_0, 0\rangle$ and $|S_1, 1\rangle \to |S_0, 0\rangle$ SE transitions. This implies that this emission stems from an essentially unrelaxed Franck-Condon region in the excited state. This is supported by observing negligible spectral intensity at lower energies, in the region of the emission spectrum (blue line). This TA spectrum undergoes a rapid temporal evolution during the first 100 fs after excitation (Fig. 2c). The strong initial SE around 2.6 eV vanishes rapidly within 50 fs, and a red-shifted SE around 2.2 eV builds up. This SE band, well reported in the literature,[14, 32, 33, 40, 49] displays no further significant spectral evolution after 100 fs, and its amplitude decays slowly on a time scale of several ps. In the high-energy region, around 2.75 eV, the long-lived GSB band remains after the initial 100-fs relaxation.

To further analyze the early-time evolution of the TA spectra, $\Delta T/T$ transients are plotted at selected probe energies in Fig. 2d. The $\Delta T/T$ signal in the low-energy part of the GSB region, at 2.64 eV (black line), shows a fast decay on a sub-50 fs time scale. The signal at 2.48 eV (red line) rapidly builds up within 10 fs and then decays within <50 fs, leaving behind a long-lived ESA signal. In the SE region, around 2.25 eV (blue line), the TA signal shows a delayed rise within <50 fs (blue line).

We globally fit[14, 67, 74] the $\Delta T/T$ spectra recorded for time delays of up to ~3 ps to quantify these observations. We obtain three exponential decays in the modelled time window (Supporting Information, section 3). The corresponding decay-associated difference spectra (DADS) are presented in Fig. 2e. As expected, the long-lived, 10.6 ps DADS component shows contributions from a GSB, a SE and an ESA band, matching earlier reports.[14, 32, 33, 40, 49] The associated decay component of 10.6 ps extends beyond the measured time window (3 ps) and therefore only reflects an effective long-time decay of the signal. Earlier measurements have shown multi-exponential decay dynamics with a component representing the slow, ns lifetime of open conformer FAD and a ~5 ps intramolecular electron transfer between the isoalloxazine and adenine moieties in stacked conformer FAD.[14, 31, 39, 40] The 0.6 ps component mainly reflects small spectral shifts and has been assigned to solvation processes.[14, 40] Interestingly, the first DADS, which has not been reported so far, has two spectral components. A positive part centered around 2.6 eV, which resembles the absorption spectrum of FAD (Fig. 2b) and reflects SE from the vertically excited Franck–Condon region. The negative part around 2.2 eV is associated with the build-up of the red-shifted SE band. The analysis suggests a short lifetime of this first component of only 20 fs. We assign this decay time to the rapid evolution of the stimulated emission spectrum immediately after photoexcitation.

**Vertical excitations from DFT/MRCI**

To confirm our assertion that the TA spectrum immediately after photoexcitation (black line in Fig. 2b) shows ESA and SE from the vertically excited (unrelaxed) Franck-Condon region, we performed DFT and subsequent DFT/MRCI (multireference configuration interaction DFT, see Methods section) simulations of riboflavin in water. Riboflavin was chosen over the full FAD to reduce computational cost, since the lowest excited states are located at the flavin part of FAD. The obtained spectra are shown in Fig. 3 and in Tables S2 & S3 in the Supporting Information. The linear absorption (Fig. 3, black line) shows two strong low-energy transitions at 2.75 and 3.37 eV as well as two weak transitions at 3.06 eV and 3.40 eV, in line with earlier reports.[47, 48, 59, 75, 76] The SE from the Franck-Condon point peaks at 2.75 eV (Fig. 3a, red line), coinciding with the $S_0 \rightarrow S_1$ transition in absorption. The SE from relaxed $S_1$ geometry is significantly red-shifted by ~400 meV (Fig. 3b, red line), as a consequence of rapid intramolecular structural reorganization on the $S_1$ potential energy surface. We also calculated the ESA spectrum (Fig. 3a, blue line) at the Franck-Condon point showing mainly three bands, in line with our experimental TA spectra (Fig. 2a). In contrast to the rapid evolution of the SE, these ESA bands are much less affected by intramolecular structural reorganization. This supports our assumption that the early time TA spectrum (black line in Fig. 2b) indeed contains the contributions of ESA and SE from the unrelaxed Franck-Condon region.

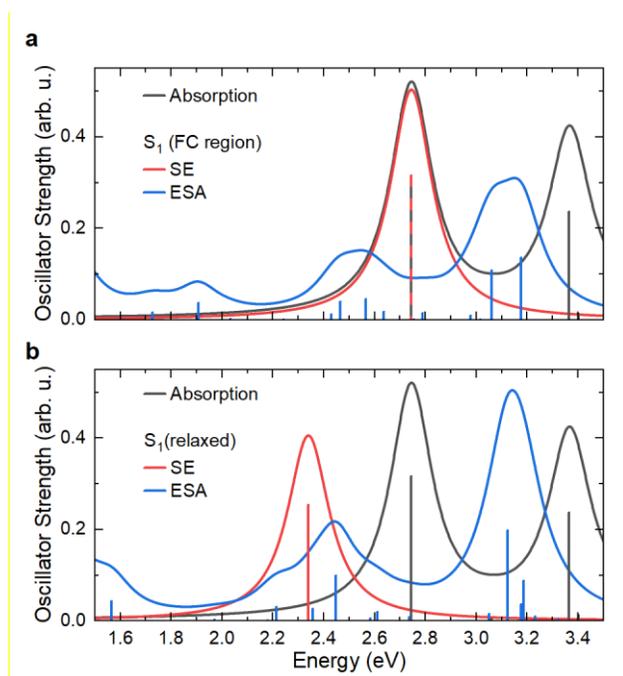

**Figure 3:** DFT/MRCI calculations; optical spectra of riboflavin. a) Linear absorption (black), SE (red, identical to the $S_1$ linear absorption peak), and ESA spectrum (blue) simulated in the unrelaxed molecular geometry at the Franck-Condon (FC) point. b) Spectra calculated from a relaxed geometry in the $S_1$ state (SE, red, and ESA, blue; linear absorption, black, as in (a) for comparison). Note the substantial red-shift of the SE band (red line). Lorentzian lineshapes with a linewidth of 200 meV (full width at half maximum)

are assumed for all transitions, as well as a scaling factor of 0.9 for all energies to improve agreement to experimental data.

**Analysis of coherent vibrational dynamics**

It is evident from Figs. 2a,d, that the transient dynamics also show pronounced oscillations that reflect the coherent vibrational motion of the molecule. In particular, high-frequency (~25 fs) modulations can be seen in the GSB region (Fig. 2d, black line), while they are essentially absent in the SE band (Fig. 2d, blue line). Slower oscillations can also be seen, especially in the region between SE and GSB. To understand the influence of coherent vibrational wavepacket motion on the excited state dynamics, we quantitatively analyze the spectral dependence of these oscillations.

To isolate these coherent modulations, we subtract the slowly varying exponential decay dynamics from the raw data (Supporting Information, section 3). In Fig. 4a, the time dynamics of the resulting residuals are exemplarily shown at two selected probe energies in the GSB (2.695 eV) and SE region (2.254 eV). In the SE region, the residuals reveal only slow oscillations with periods of 50 to 120 fs. In the GSB region, instead, high-frequency oscillations with periods down to 20 fs are visible. To reveal their vibrational frequencies, we perform a Fourier transform (FT) analysis of these residuals along the delay axis. The resulting FT spectra are shown in Fig. 4b. Below 1000 $cm^{-1}$, we observe the same dominant low-frequency modes in both spectra. High-frequency modes in the 1100-1700 $cm^{-1}$ range are present in the GSB region but are almost absent in the SE region. To correlate the oscillation amplitudes of these modes with the different contributions to the TA signal (Fig. 4c), we show the relevant GSB, SE and ESA components in Fig. 3c and plot a map of the FT spectra as a function of probe energy in Fig. 4d. A FT spectrum that is integrated along the probe energy axis with ~10 $cm^{-1}$ resolution is shown in Fig. 4e. The resulting mode frequencies are in quantitative agreement with reported Raman spectra. A direct comparison with literature values is shown in Table S1 of the Supporting Information.[77-79] Importantly, the FT map clearly reveals that the high-frequency modes are primarily localized in the GSB region, centered around the 0-0 transition close to 2.6 eV. Their amplitude vanishes in the SE region. The probe energy dependence of all high-frequency modes is very similar, as shown in Figs S9 and S10 in the Supporting Information. In stark contrast, the low-frequency modes display more diverse spectral profiles. Generally, these profiles, discussed in more detail in Fig. 5, are shifted towards lower probe energies and overlap with the SE and ESA regions. Additionally, some of the low-frequency modes, in particular 285 cm$^{-1}$, 424 cm$^{-1}$, and 607 cm$^{-1}$, show substantially larger amplitudes than all high-frequency modes.

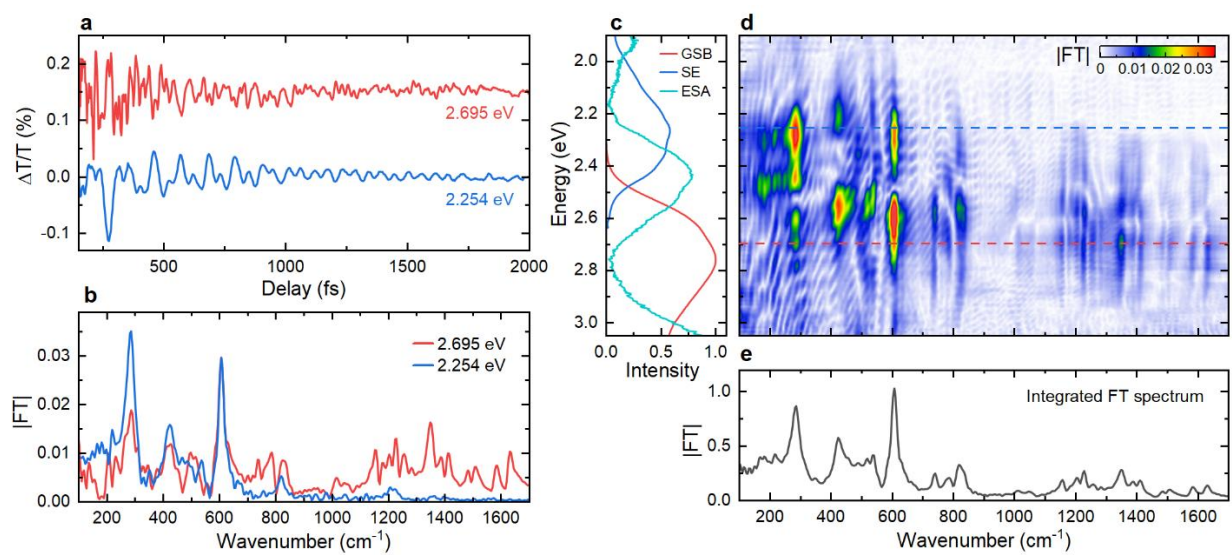

**Figure 4:** Fourier transform (FT) analysis of coherent vibrational dynamics in FAD dissolved in water. (a) Residuals from the GSB region (2.695 eV) and the SE region (2.254 eV) were obtained by subtracting time-dependent DADS from the experimental $\Delta T/T$ data. The residuals have been vertically shifted for clarity. (b) Comparison of the FT spectra deduced from the residuals in (a). High-frequency components (>1000 cm$^{-1}$) are absent in the SE spectra (blue), while the dominant low-frequency modes show similar amplitudes. (c) Decomposition of the quasi-static DADS (blue line in Fig. 2e, not shown here) into GSB (red), SE

(blue), and ESA (cyan) components, assuming that the line shapes of the GSB and SE components match those observed in linear spectroscopy (Fig. 1). (d) Two-dimensional map of the FT amplitude as a function of mode frequency and probe energy. The dashed lines mark the FT spectra plotted in (b). (e) FT spectrum obtained by integrating the map along the probe energy. The integrated spectrum contains all coherently excited ground and excited state modes and is in good agreement with reported Raman spectra of FAD in water.

We want to argue that these differences arise because the high-frequency oscillations correspond to wavepacket motion in the $S_0$ electronic ground state, while low-frequency oscillations are triggered in the excited $S_1$ state. For this, we report the probe energy dependence of the amplitude and phase of the complex Fourier spectra of several selected modes in Fig. 5. These data are shown for two selected low-frequency modes, 424 cm$^{-1}$ (b) and 607 cm$^{-1}$ (c), and two high-frequency modes at 1228 cm$^{-1}$ (d) and 1349 cm$^{-1}$ (e). The amplitude profiles of both high-frequency modes are centered in the GSB region and exhibit two distinct minima at 2.48 eV and 2.63 eV, each coinciding with a $\pi$ phase jump. No significant amplitude is observed in the SE region between 2.0 and 2.3 eV. In contrast, the low-frequency modes exhibit substantial amplitude in both the SE region and the GSB region. The phase profiles of the low-frequency modes show differences. The 607 cm$^{-1}$ mode displays a phase jump from $-\pi$ to 0 around the amplitude dip at 2.48 eV. In contrast, the 424 cm$^{-1}$ mode undergoes a phase jump from 0 to $-\pi$ around 2.4 eV.

To rationalize these observations, it is instructive to recall detailed experimental[33, 49] and theoretical studies[47] of the ultrafast coherent dynamics of flavin molecules. Weigel et al.[49] have used TA spectroscopy with 30-fs pulses and 50-fs pump-probe cross correlations to study such dynamics. They observed the SE band already during their time resolution and have assigned its formation to an ultrafast (< 20 fs) $\pi\pi^* - n\pi^*$ vibronic coupling process. In later work, Weigel et al.[33] applied Femtosecond Stimulated Raman Spectroscopy to access the vibrational modes in the ground $S_0$ and excited $S_1$ state of riboflavin and FAD in different solvents. The reported Raman bands have been analyzed and assigned using a quantum-mechanical normal mode analysis. Generally, the observed mode frequencies are in very good agreement with our observations (Table S1 in the Supporting Information). Also in this work, time-resolved fluorescence and TA spectra recorded for time delays of more than 100 fs reveal the presence of a SE around 2.25 eV during the experimental time resolution. In a theoretical study by Klaumünzer et al.[47] the nonadiabatic dynamics of riboflavin in the gas phase and in a microsolvated water environment were calculated using a mixed quantum classical surface hopping approach. Their calculations show the appearance of a distinct, high-energy SE band reflecting emission from the Franck-Condon region shortly after photoexcitation. The lineshape of this band is similar to the linear absorption spectrum and shows vibronic substructure resulting from the coupling to C-C or C-N stretching modes with 22-fs period. This high-energy SE is present for no more than 10 fs. The excited state wavepacket motion that is associated with this mode decays rapidly within < 50 fs, resulting in the build-up of a strongly red-shifted SE band. This red-shift is more pronounced in the microsolvated environment. These ultrafast dynamics reflect a very rapid IVR process in riboflavin that results from vibronic coupling of the bright $\pi\pi^*$ and the dark $n\pi^*$ state that is induced by these high-frequency vibrational modes. Since the optical properties of all flavins are largely governed by those of their isoalloxazine ring, we expect that the results of these calculations are directly transferable to FAD.

In accordance with our simulations shown in Fig. 3, this suggests that the strong $\Delta T/T$ band around 2.65 eV that is seen in Fig. 2b indeed reflects SE from the Franck-Condon region immediately after photoexcitation. The pronounced red-shift that is seen in Fig. 2a (black arrow) is now readily understood. It likely reflects an IVR process in FAD which is exceptionally fast due the strong nonadiabatic coupling in flavins. This IVR process seems to be complete within 50 fs since we do not observe significant changes in the red-shifted SE spectrum after this redistribution process. This rapid IVR process also explains the absence of high-frequency vibrational oscillations in the red-shifted SE region that we observe in Fig. 4b,d since it quickly dampens high-frequency vibrational motion in the excited state. The low-frequency modes, instead, appear unaffected by this IVR process.

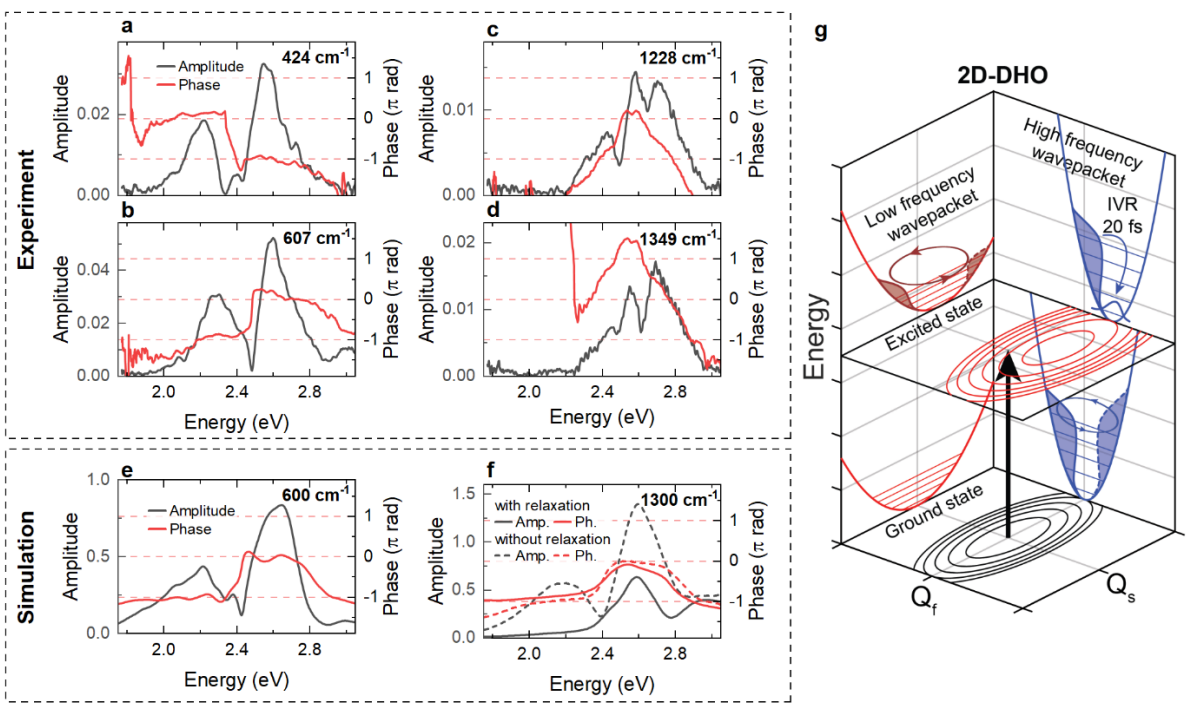

**Figure 5:** Probe energy dependence of the FT analysis in Fig. 4 for selected high- and low-frequency modes of FAD. (a-d) Amplitude (black) and spectral phase profiles (red) for the 424 cm$^{-1}$ (a), 607 cm$^{-1}$ (b), 1228 cm$^{-1}$ (c), and 1349 cm$^{-1}$ (d) modes obtained by FT analysis of the coherent $\Delta T/T$ oscillations. (e-f) Amplitude (black) and phase profiles (red curve) for a 600 cm$^{-1}$ (e) and 1300 cm$^{-1}$ (f) mode simulated using a two-dimensional displaced harmonic oscillator (2D-DHO) model. For the high-frequency mode (f), simulations in the absence (dashed) and presence (solid) of rapid, 20-fs excited state vibrational relaxation are reported, whereas relaxation is neglected for the low-frequency mode. With relaxation, the amplitude

vanishes in the SE region, in good agreement with the experimental results in panel (b) and (d). (g) Sketch of the relaxation scenario in a 2D-DHO model. Impulsive optical excitation launches a vibrational wavepacket in the excited-state Franck-Condon region (black arrow). IVR leads to rapid relaxation along the high-frequency $Q_f$ coordinate, while coherent oscillations persist along $Q_s$. In contrast, ground state wavepacket oscillations are launched along $Q_f$, but are much weaker along $Q_s$.

**Displaced harmonic oscillator simulations**

To test how such rapid IVR manifests in our TA spectra, we set up the conceptually simplest phenomenological model that can account not only for the transient spectral shifts but also for the coherent oscillations in our spectra. We choose a model of two harmonic oscillators that are coupled to an electronic two-level system, as sketched in Fig. 5g. We simulate TA signals based on this 2-mode DHO model by non-perturbatively solving the Lindblad master equation for the density matrix of our system[72] to account for vibrational energy relaxation in the $S_1$ state and for pure electronic dephasing. We simulate the effects of ground and excited state vibrational dynamics launched by a 12-fs pulse centered around 2.64 eV. To account for the linear absorption spectrum of FAD, we consider a single, high-frequency mode with an effective frequency of $\omega_{vib,1} = 160$ meV (~1300 cm$^{-1}$) and dimensionless displacement of $\Delta_1 = 1.6$. For the second mode, we choose $\omega_{vib,2} = 74$ meV (~600 cm$^{-1}$) and $\Delta_2 = 1.2$ (Supporting Information, section 5). With this, we can indeed qualitatively reproduce the experimental absorption (Fig. S12) and emission spectrum. Obviously, our simplified DHO model does not capture the full Stokes shift that is seen in the experiment since it neglects the structural reorganization in the $S_1$ state that is demonstrated by the vertical excitation calculations in Fig. 3. When simulating the TA dynamics, we mimic the complex multimode IVR process, resulting in a rapid loss of vibrational coherence, by introducing a finite vibrational relaxation time of 20 fs using an appropriate Lindblad operator for the high-frequency mode in the electronically excited state.

In such a DHO model, coherent vibrational wavepackets are launched in the excited state after impulsive excitations via a vertical Franck-Condon transition.[80, 81] The wavepacket motion can be read out via SE transitions showing periodic modulations with the frequencies of the two modes and wavepacket oscillation amplitudes that increase with mode displacement.[82] Ground state vibrational wavepackets are created via impulsive stimulated Raman scattering[83] and read out via GSB transitions.[82] For short excitation pulses, the wavepacket oscillation amplitude decreases linearly with pulse duration since a propagation of the excited state wavepacket is needed to reach a finite displacement in the ground state.[81] For our 12 fs pump pulses, this implies that wavepackets along both vibrational coordinates are launched in the excited state (Fig. 5g). In contrast, ground state wavepacket motion is predominantly excited along the high-frequency coordinate while the displacement of the launched low-frequency wavepacket is much weaker. TA maps resulting from these 2D-DHO simulations are shown in Figs S13 and S14. Importantly, they show a SE spectrum with a shape that mirrors the absorption and displays a Stokes shift of ~400 meV (Fig. S12). This Stokes shift is significantly smaller than the experimental one observed in Fig. 1b since intramolecular structural relaxation (Fig. 3) is neglected.

The simulated TA maps show persistent oscillations at the frequencies of both vibrational modes. An analysis of the spectral mode profiles of the Fourier transforms of these oscillations is particularly instructive. Such a mode profile is shown for the low-frequency mode in Fig. 5e. It is essentially independent of the relaxation time of the excited state high-frequency mode. In agreement with experiment (Fig. 5b), it shows large amplitude in both the GSB (2.6 eV) and SE (2.3 eV) regions. The amplitude dip at ~2.4 eV coincides with a phase jump from $-\pi$ to 0. As in experiment, the amplitude abruptly drops for energies above 2.8 eV, i.e., in a spectral region where SE is strongly reduced. This is the signature that the low-frequency wavepacket is predominantly launched in the excited state and is read out via SE. While this DHO model can reproduce the amplitude and phase profiles for the 607 cm$^{-1}$ mode, the 424 cm$^{-1}$ mode shows qualitatively different behavior, in particular the different sign of the phase jump. This suggests that here, the excited state wavepacket is not read out via SE but rather via an ESA to a higher-lying electronic state. Similar phase signatures are also seen for several other low-frequency modes (see Fig. S9 and S10). Since, so far, the knowledge about the relevant higher-lying potential energy surfaces is limited, we did not attempt a quantitative modelling.

In the absence of vibrational relaxation, the spectral mode profile of the 1300 cm$^{-1}$ mode is rather similar to that of the 600 cm$^{-1}$ mode, except for a significantly larger amplitude at energies above 2.8 eV, i.e., on the high-energy side of the GSB band (Fig. 5f, dashed line). The reason for this difference becomes evident when vibrational relaxation is turned on (Fig. 5f, solid line). Now, the spectral amplitude in the SE region below 2.3 eV vanishes, and the amplitude between 2.4 and 2.8 eV is reduced by half. This difference in amplitude is readily understood. Now, the SE contribution of the excited state wavepacket to the spectral profile is absent. Above 2.8 eV, however, the SE amplitude is negligible, and the probe pulse can only sense the ground state wavepacket that is launched by stimulated impulse Raman scattering. Therefore, the amplitude in this high-energy region remains unchanged when altering the relaxation rate for the high-frequency mode, since only ground state wavepacket motion is probed.

Despite the simplicity of the 2D-DHO model, the simulated spectral mode profiles in the presence of rapid vibrational relaxation are in reasonable agreement with experiment. In contrast, those calculated without relaxation show a high amplitude in the 2.3-eV SE region, which is not observed in the experimental data.

**Summary and conclusion**

In summary, we have observed, immediately after impulsive photoexcitation of FAD molecules in aqueous solution, a short-lived stimulated emission band from the Franck-Condon region of the electronically excited $S_1$ state. This band decays within 50 fs, resulting in the build-up of a substantially red-shifted, relaxed stimulated emission band. Our data show that this rapid relaxation process is concurrent with an essentially complete damping of coherent oscillations of all high-frequency (>1000 cm$^{-1}$) modes in the electronically excited state, while low-frequency, excited-state wavepacket motion persists for up to 1 ps. Our results align well with quantum-dynamical simulations for riboflavins, which support the assignment of our TA spectra and relate this mode-selective and fast intramolecular vibrational energy redistribution process to nonadiabatic couplings between the optically bright $S_1$ state and an energetically close-lying dark state,[47] mediated by the high-frequency carbon backbone modes of the isoalloxazine moiety.

As such, our results may be of direct relevance for closely related FAD-binding cryptochrome proteins for which nonadiabatic couplings between $S_1$ and $S_2$ have recently been predicted theoretically.[50] This raises the question which role such nonadiabatic couplings play for the dynamics and yield of charge transfer processes[14] and radical pair formation[21] in these magnetically-sensitive proteins. Experiments in this direction are currently underway in our laboratory. More generally, our results present an example for very rapid IVR processes driven by nonadiabatic couplings that are caused by (a multitude of) high-frequency vibrational modes. It appears to us that the quantum dynamics of such processes and their functional relevance for charge-separation processes in molecules and nanosystems are currently understood only to a limited extent, at least experimentally.[53] Pump-probe and, in particular, two-dimensional electronic spectroscopy with few-fs or even sub-fs time resolution may provide desirable new insight.

**Methods**

**Ultrafast pump-probe spectroscopy**

Ultrafast pump-probe experiments are conducted using a home-built setup based on hollow-core fiber supercontinua (see Fig. S1).[63] A 1-m hollow-core fiber (Savanna, Ultrafast Innovations) filled with 2.4 bar Neon gas (absolute pressure) is pumped by a regenerative Ti:sapphire laser (Legend Elite, Coherent) that outputs 26-fs pulses with 1 mJ pulse energy centered at 800 nm with 10 kHz repetition rate. The generated supercontinuum spectrum, spanning ~350-1000 nm,[63] is filtered to a range of ~390-720 nm and split into a pump and probe beam. While the probe is used directly, the pump is further spectrally filtered in the spectral domain using a 4f setup. Pump spectra are set to cover a range from ~435-500 nm using a tunable slit in the Fourier plane of the 4f setup. Chirped mirrors (DCM12, Laser Quantum) are used to compensate for dispersion of pump and probe. This yields a pump pulse duration of 12 fs as measured via transient grating frequency-resolved optical gating (see Fig. S2). The delay between pump and probe $t_d$ is set via a retroreflector mounted on a motorized translation stage (M112.1DG, Physik Instrumente) and both beams are focused into the 1 mm quartz sample cuvette using an off-axis parabolic mirror to ~40 μm spot size. The transmitted probe beam is spectrally dispersed and spectra $S_{pu,pr}(\lambda)$ are recorded with full laser repetition rate using a fast and sensitive line camera (Aviiva EM4, e2v). Mechanical chopping of pump and probe using a custom chopper blade with 2:1 duty cycle allows to record scattering-corrected differential transmission spectra

$$\frac{\Delta T}{T}(t_d, \lambda) = \frac{S_{on,on} - S_{off,on} - S_{on,off}}{S_{off,on}} \quad (1)$$

from all combinations of blocked (off) and transmitted (on) pump and probe.[63] Polarizations of pump and probe are set via thin-film polarizers.

For all experiments, flavin adenine dinucleotide (FAD) as purchased from Sigma Aldrich is dissolved in water with a concentration of 200 µM and experiments are performed at room temperature. For each dataset, a solvent reference measurement of bare water is performed under identical experimental conditions. The experimental data recorded with compressed probe pulses is shown in Fig. 2. To avoid any spurious effects of higher-order phases resulting from the probe compression, we additionally record TA data with an uncompressed, chirped probe (shown in Figs. 4 and 5) for quantitatively analyzing spectral mode profiles.

For the first dataset recorded with compressed probe, the relative polarization between pump and probe is set to the magic angle (54.7°) and a pump pulse energy of 25 nJ is used. For the second dataset with chirped probe, the relative polarization is tuned parallel and a pump pulse energy of 20 nJ is used. Pump-probe data is recorded with 5 fs step size up to a delay of 3 ps. More details can be found in the Supporting Information section 1 and 2.

**Data evaluation**

For both datasets, pump-probe transients of FAD and the water reference contain strong cross-phase modulation[70] (XPM) signal centered around zero delay. This XPM is used to determine the wavelength-dependent time zero arising from the chirp of the probe pulse and all measurements are shifted accordingly. Using the solvent reference, the XPM contribution is carefully subtracted from the FAD data (see Supporting Information section 3). A global analysis is performed on the data using a Matlab-based toolbox,[14, 67, 74] yielding decay-associated difference spectra. For analyzing the coherent vibrational dynamics, the fit obtained from the global analysis is subtracted from the data to obtain residuals. A Fourier transform (FT) of these residuals yields (complex) FT maps as a function of the vibrational frequency and probe energy. Additionally, spectral mode profiles in amplitude and phase can be obtained for each vibrational mode by taking crosscuts through the FT map. For more details, see Supporting Information section 3.

**DFT calculations**

The riboflavin model was built starting from the protein data bank entry for FAD, conveniently labeled "FAD", and then removing the structure to riboflavin by replacing the first phosphorus atom with hydrogen and deleting all subsequent atoms in the structure. The structure was then relaxed using the CAM-B3LYP[84] functional and the def2-TZVP[85] basis set using the ORCA[86] software, employing an implicit water solvation (CPCM).[87] The most recent version of the DFT multireference configuration interaction (DFT/MRCI)[88] approach, the R2022 Hamiltonian, was then used to obtain the spectra depicted in Fig. 3, based on a BHLYP[89] orbital guess with def2-SVP basis. Energies were found to be slightly overestimated; we thus employed a scaling of 0.9 to improve agreement with experimental values.

For relaxed riboflavin (TD-DFT[90] optimization of the $S_1$ state, in CPCM), the same DFT/MRCI procedure (see previous paragraph) was applied; the optimization started from the Franck-Condon point.

## Supporting Information

Experimental details, data evaluation, vibrational mode profiles, comparison with literature, simulation details, additional simulation results, Figs. S1-S14, Tables S1-S4.

## Acknowledgements

We acknowledge funding from Deutsche Forschungsgemeinschaft (SFB 1372 "Magnetoreception and navigation in vertebrates" (project number 395940726), Li 580/16-1, DE 3578/3-1, JPG GO 2430/3-3) and the Niedersächsische Ministerium für Wissenschaft und Kultur (DyNano and Wissenschaftsraum ElLiKo).

## Data availability

The experimental data supporting the claims in this study are presented in the manuscript and in the Supporting Information in graphic form and can be obtained from the authors upon reasonable request.

## Notes

The authors declare no competing interests.